# Polaron relaxation in ferroelectric thin films

E.Arveux, D.Levasseur, A.Rousseau, S.Payan, M.Maglione

ICMCB-CNRS, Université de Bordeaux, 87 av Dr Schweitzer 33608 Pessac France

G.Guegan

STMicroelectronics, 16 Rue Pierre et Marie Curie 37071 Tours France

Abstract. We report a dielectric relaxation in ferroelectric thin films of the $ABO_3$ family. We have compared films of different compositions with several growth modes: sputtering (with and without magnetron) and sol-gel. The relaxation was observed at cryogenic temperature (T<100K) for frequencies from 100Hz up to 10MHz. This relaxation activation energy is always lower than 200meV. It is very similar to the polaron relaxation that we reported in the parent bulk perovskites. Being independent of the materials size, morphology and texture, this relaxation can be a useful probe of defects in actual integrated capacitors with no need for specific shaping.

From the very beginning, the occurrence of charged defects has been recognized as a key issue for the optimization and long term use of ferroelectric films [1]. The low oxygen partial pressure during the vacuum films deposition, the cationic segregation in polycrystalline films and the oxygen diffusion during the films operation under electrical stress are among the possible sources of such charged defects. Deep investigation of charged defects is underway including ex-situ and in-situ XPS,RBS/NRA measurements [2-4]. While the physics of defects in ferroelectrics films has some specific features, it shares a lot of trends with related investigations in the parent bulk ferroelectric perovskites [5]. In single crystals, by combining ESR spectroscopy and IR absorption [6-7], clear evidence of polaronic state was observed, which is a highly probable outcome of oxygen vacancies and other charged defects. We have previously shown that such polarons lead to a small but well defined dielectric relaxation in 100 different bulk perovskites of different compositions, morphologies and oxidation states [8]. Occurring always below 100K in the sub-MHz frequency range, this relaxation is thermally activated with an activation energy lying in the 100meV range.

In the present report, we show that such polaronic relaxation also occurs in thin films of different composition ($BaTiO_3$, $(Ba,Sr)TiO_3$ and $Ba(Ti,Nb)O_3$) obtained with several growth modes: sputtering (with and without magnetron) and sol-gel. This relaxation exhibits the same features as in the parent bulk compounds. Not only this result confirms the very microscopic origin of this polaronic relaxation but it also establishes the use of low temperature dielectric spectroscopy as a simple tool for a qualitative classification of integrated ferroelectric capacitors with no need of specific shaping.

The ferroelectric films were grown on Platinized Silicon substrates of diverse origin. The first set of films was sputtered from home made targets using 4 different deposition chambers corresponding to 4 types of target/substrate assemblies: on- and off-axis, substrate above or below the target. The second set of films was grown using sol gel spinning on platinized silicon wafers. In addition to the compositional changes, the investigated films are very diverse: while sputtered films are textured with several preferred orientation, sol gel films have random orientation. The columns diameter in sputtered films and the grain size in sol gel films lie in the sub-100nm range. Since the processing and post-processing annealing took place at different temperatures and different atmospheres, the density of oxygen vacancies can also change a lot among all these films. Even though a quantitative estimate of the actual density of these oxygen vacancies is out of reach, we will show later that it may play a decisive role in our low temperature dielectric observations.

These dielectric experiments were all carried out in the Metal-Insulator-Metal (MIM) geometry. The sample temperature could be monitored from 4K to 340K with an accuracy better than 0.1K using a quantum design dewar. The operating frequency has been swept from 100 Hz to 10 MHz. A typical variation of the capacitance and of the dielectric losses is shown on figure 1 for selected spot frequencies and for the whole temperature range for a sputtered $BaTiO_3$ film. On this figure, two anomalies are clearly evidenced at about 250K and 50K.

We will first focus on the high temperature feature which was already reported in several papers [9-10] for $BaTiO_3$ films with different electrodes. The maximum in losses can be ascribed to a dielectric relaxation whose activation energy can be computed from an Arrhenius law linking the operating frequency $f$ to the related temperature T where the loss maximum occurs $f = f_0 e^{-\frac{E}{kT}}$. The activation energy for this high temperature relaxation is about 100meV in all cases. However, this relaxation is not observed in all films depending on the growth conditions. We ascribe this high temperature relaxation to space charge localization at electrodes/film interfaces [11] and we exclude any intrinsic origin for it [10]. Because of its large variability, this interface related relaxation will not be discussed further.

We rather focus now on the low temperature dielectric relaxation recorded in all of the 15 investigated films. Typical loss maximums are shown on figure 2 for sputtered (figure 2a) and sol gel (figure 2b) BST at the frequency range of 10 kHz-10MHz. As for the high temperature relaxation, pointing the temperature at which the maximum occurs for a given frequency leads to a set of (f,T) couples which can be plotted in an Arrhenius f vs 1/T using a semi-logarithmic scale.

Arrhenius plot together with their fit with equation $f = f_0 e^{-\frac{E}{kT}}$ are displayed on figure 4 where $f_0$ is the high temperature extrapolation of the relaxation frequency $f$, E its activation energy and k the Boltzmann constant. We first underline that the Arrhenius law holds for all of the investigated films and that the activation energy is always low E<150meV. The extrapolated high temperature frequency $f_0$ is very much fluctuating because of the limited and low temperature range where the relaxation takes place. We will thus now discuss the computed activation energies and their possible link with the film processing, structure and content.

In figure 3 inset, one can see that these activation energies are of the same order (<200meV) as the one already reported in a large number of bulk perovskites whatever their morphology (ceramics or single crystals), composition, doping and ferroelectric properties [8]. Such persistence called for a common origin which was ascribed to polaron-related dipolar relaxations. These polarons can originate from charged point defects which are always present in oxide perovskites, even the purest single crystals in which residual Fe impurities or oxygen vacancies remain. Even though such polaron model was heavily debated [12-14], we can transfer it to our thin films since the atomic density of oxygen vacancies is always high in vacuum deposited sample as well as in sol-gel processed films. Within this general model, we try now to sort the activation energies that we found versus the processing conditions keeping in mind that the key for the polaron relaxation to occur is the stabilization of charged point defects having several valencies:

-the lowest activation energies are found for samples $BaTiO_3$ (56, 20) which were room temperature sputtered using a standard sputtering. ESR experiments previously shown that such columnar films contain a lot of oxygen vacancy-related defects like $Ti^{3+}$-VO [15]; when a bias was applied to the substrate during the growth, the density of such defects was shown to be the highest and this is the case of BTB20 which shows the smallest activation energy of 10meV (sample 20 not shown on figure 3). For such films, we ascribe the extremely low activation energy to the easy hoping of polarons among very closely spaced charged defects.

-intermediate activation energies occur in magnetron sputtered films which were high temperature processed either during the growth or post-annealed. In these $BaTiO_3$ and BST films (10,325, 326,339,349) the texture degree could be monitored between textured and fully disoriented [to be published]; the polaron activation energy is increasing with the disorientation degree of the films from about 80meV for BT325 up to 140meV for BT365 with several intermediate cases. As for the density of charged defects, work is in progress to get a comparative estimation with previous films. We only can state that these magnetron sputtered films are closer to stoichiometry than the non-magnetron sputtered film mentioned above.

-for sol gel BST films (103, 106) the activation energy is always in the high range which is in between 130 and 180meV; this low hopping probability of polarons may be related to the low density of charged defects in such optimized films .

To summarize, we have shown that a small amplitude dielectric relaxation is always observed at cryogenic temperature in ferroelectric perovskite thin films. This observation is very similar to what was reported in a large number of bulk materials of the same family. The relaxation occurs in all films whatever their morphology, texture, chemical content and processing way. We suggested a very preliminary model trying to map out the relaxation activation energies with the amount of charged defects. Further investigations are needed to better quantify such a link. Low temperature dielectric spectroscopy might appear as a simple tool –albeit non quantitative- for classifying integrated ferroelectric capacitors with no need for a specific shaping or preparation prior to the investigation. This could be helpful in the process of optimization of ferroelectric films for decreasing their dielectric losses and leakage current.

Acknowledgements

This work was supported by the European Multifunctional Materials Institute (EMMI), the Conseil Régional d'Aquitaine and the French National Agency for Research under the project ANR ABSYS. We also thank Rodolphe Decourt who has built a very suitable measurement setup for cryogenic temperature dielectric experiments on thin films. The solutions used for the processing of sol-gel films were provided by Mitsubishi Corp. Japan.

Figure Captions

Figure 1: capacitance (left scale) and dielectric losses (right scale) of a sputtered $BaTiO_3$:Nb film showing interface relaxation at high temperature and a polaronic relaxation for lower temperature (T<100).

Figure 2: low temperature relaxation of BST film (a) sputtered deposited   and (b) sol-gel deposited. Spread of the loss peak over temperature is larger in the former case showing lower activation energy.

Figure 3: relaxation map for several of the investigated films:  SS56 (standard sputtered $BaTiO_3$ without substrate heating); MS10, MS325, MS326, MS339, MS349 (magnetron sputtered $BaTiO_3$ and BST with substrate heating); SG103,SG109 (sol gel deposited BST). From Arrhenius fitting, the line slopes give the activation energies (inset) which are the highest for sol gel optimized films and the smallest for standard sputtering, this latter having a large density of point defects.

Figure 1

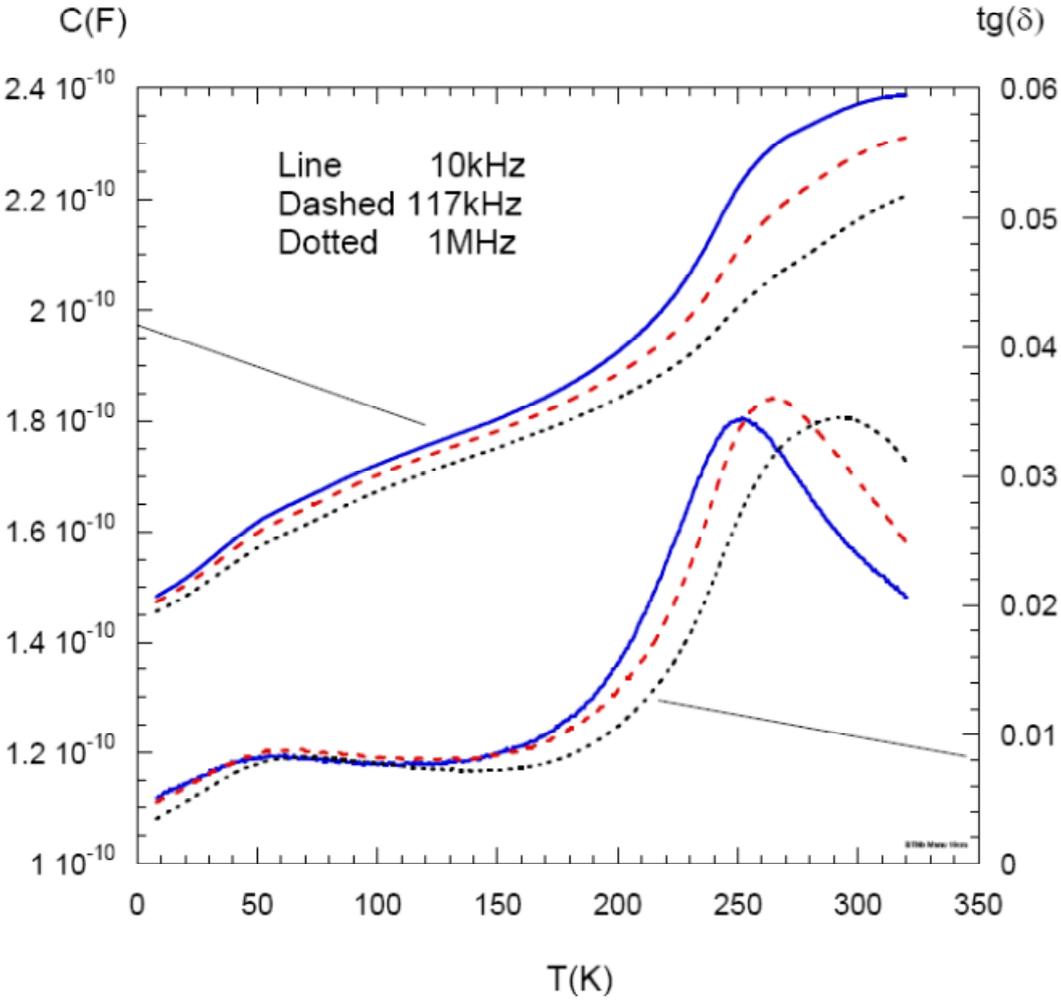

Figure 2a

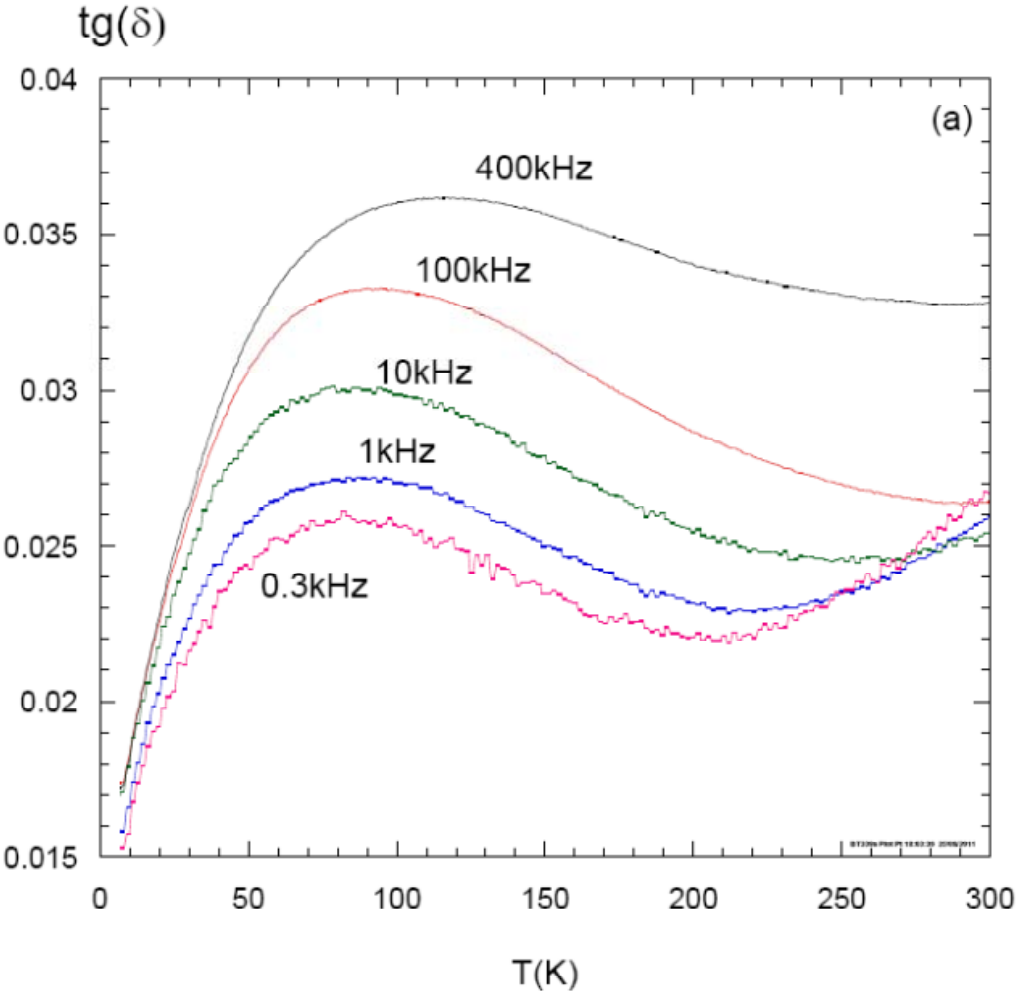

Figure 2b

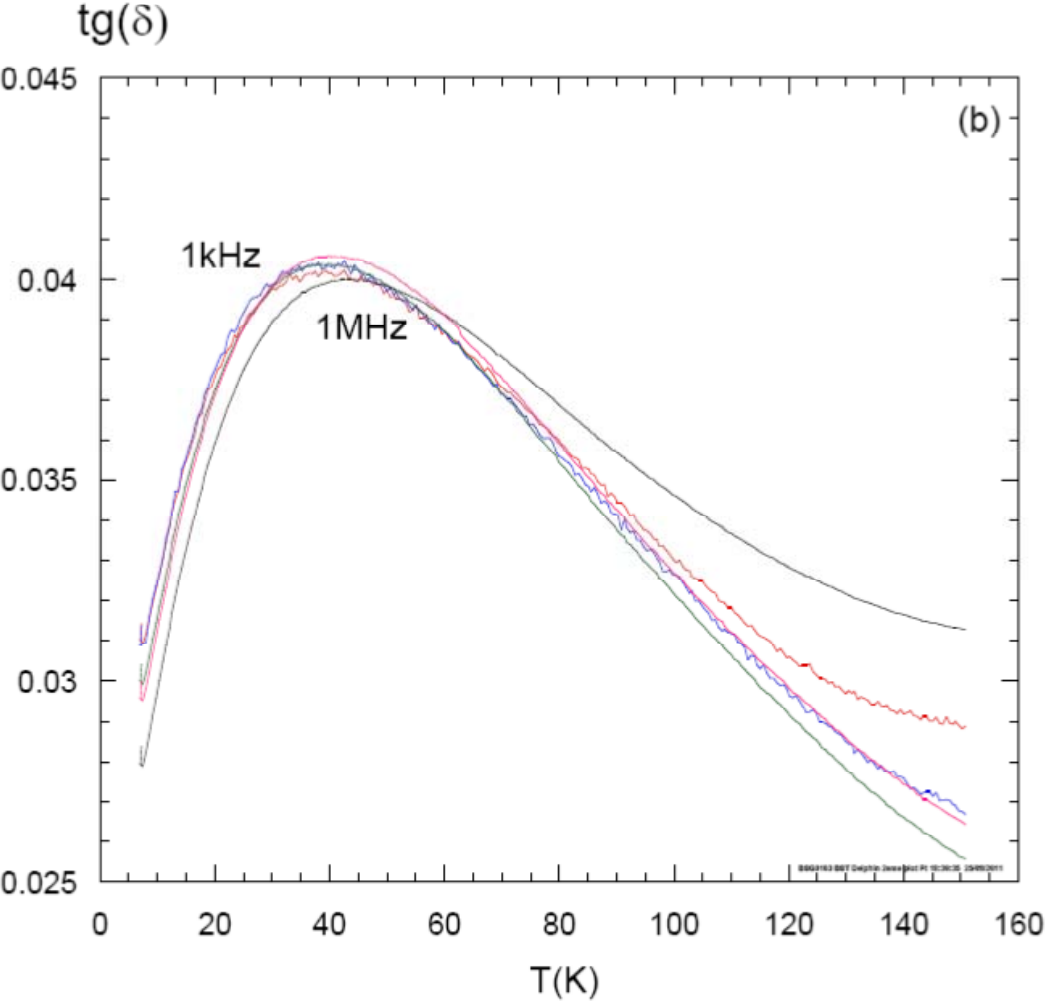

Figure 3

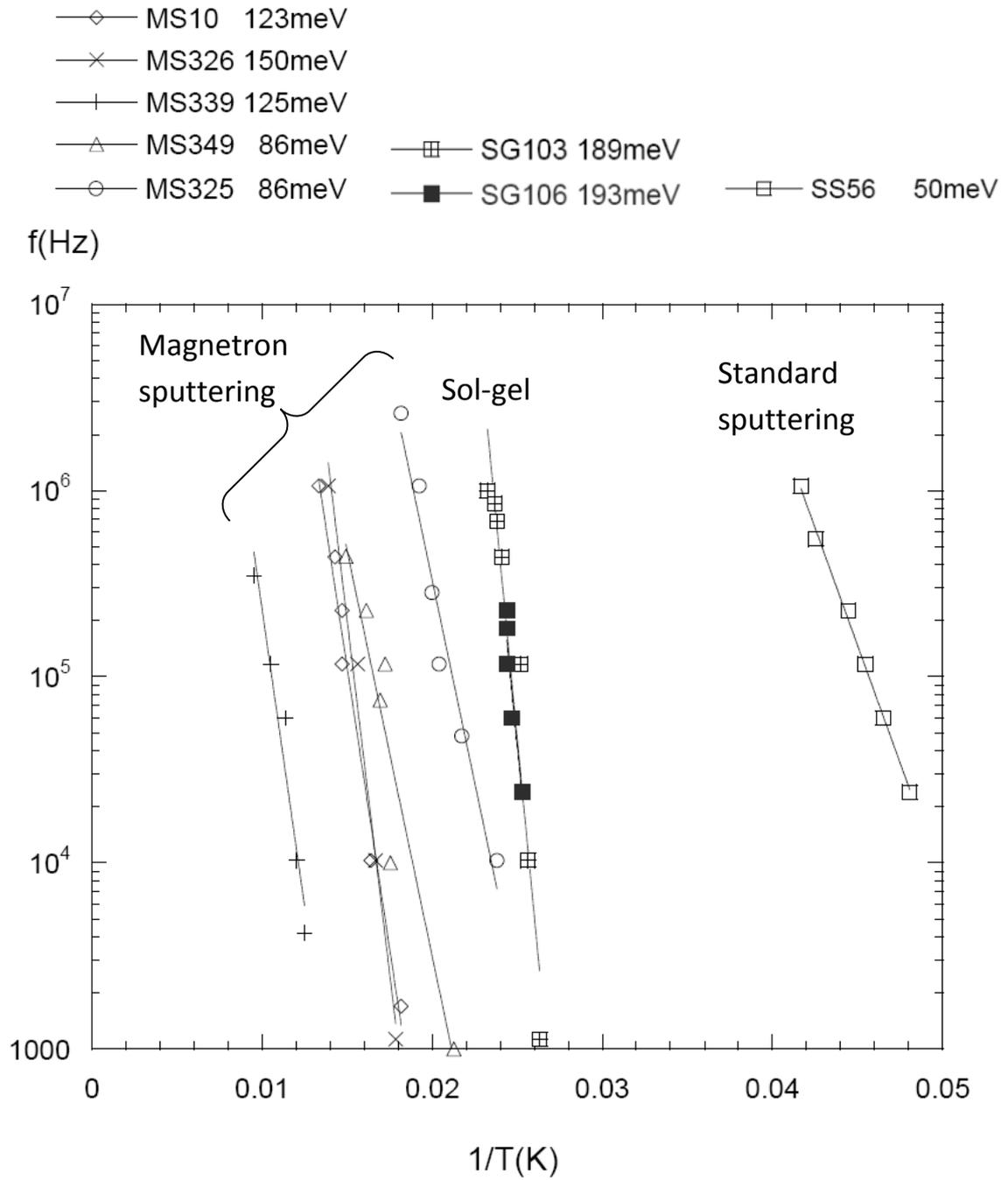